# Progress toward a first observation of parity violation in chiral molecules by high-resolution laser spectroscopy

Dedicated to the memory of Professor André Collet

B. Darquié[1,*], C. Stoeffler[1], A. Shelkovnikov[1], C. Daussy[1], A. Amy-Klein[1], C. Chardonnet[1], S. Zrig[2], L. Guy[3], J. Crassous[2,*], P. Soulard[4], P. Asselin[4], T. R. Huet[5], P. Schwerdtfeger[6], R. Bast[7], T. Saue[8]

[1] *Laboratoire de Physique des Lasers, UMR7538 Université Paris 13-CNRS, 99 av. J.-B. Clément, F-93430 Villetaneuse, France*
[2] *Sciences Chimiques de Rennes, Campus de Beaulieu, UMR6226 Université de Rennes 1-CNRS, 35042 Rennes Cedex, France*
[3] *Laboratoire de Chimie, UMR 5182 Ecole Normale Supérieure de Lyon-CNRS, 46 Allée d'Italie, F-69364 Lyon 07, France*
[4] *Laboratoire de Dynamique, Interactions et Réactivité, UMR7075 Université Pierre et Marie Curie-CNRS, 4 Place Jussieu, F-75005 Paris, France*
[5] *Laboratoire de Physique des Lasers, Atomes et Molécules, UMR8523 Université Lille 1-CNRS, F-59655 Villeneuve d'Ascq Cedex, France*
[6] *Centre for Theoretical Chemistry and Physics, New Zealand Institute for Advanced Study, Massey University Albany, Private Bag 102904, North Shore City, Auckland 0745, New Zealand*
[7] *Centre for Theoretical and Computational Chemistry, Department of Chemistry, University of Tromsø, N-9037 Tromsø, Norway*
[8] *Institut de Chimie, UMR7177 Université Louis Pasteur-CNRS, 4 rue Blaise Pascal, F-67000 Strasbourg, France*

* benoit.darquie@univ-paris13.fr, jeanne.crassous@univ-rennes1.fr

## Abstract

Parity violation (PV) effects in chiral molecules have so far never been experimentally observed. To take this challenge up, a consortium of physicists, chemists, theoreticians and spectroscopists has been established and aims at measuring PV energy differences between two enantiomers by using high-resolution laser spectroscopy. In this article, we present our common strategy to reach this goal, the progress accomplished in the diverse areas, and point out directions for future PV observations. The work of André Collet on bromochlorofluoromethane (**1**) enantiomers, their synthesis and their chiral recognition by cryptophanes made feasible the first generation of experiments presented in this paper.

## I. Introduction

Chirality is a fundamental concept in chemistry, biology and physics. For chemistry chirality is a challenge, notably in the development and synthesis of molecules for pharmaceuticals, agrochemicals, flavors and, more recently, in supramolecular chemistry and nanotechnology[1-3]. The development of catalysts for stereoselective synthesis is one of the most important tasks of modern chemistry. In biology, homochirality is a hallmark of life in that nature shows, with very few exceptions, a distinct preference for *L*-amino acids and *D*-sugars over their mirror images. However, the origin of biohomochirality remains unknown[4-6]. In physics, objects of opposite chirality are connected by the parity operation (P), which combines with charge conjugation (C) and time reversal (T) in the CPT theorem: a mirror universe where, all particle positions are reflected about some plane (parity inversion), all particles are replaced by their anti-particles (charge conjugation) and all momenta are reversed (time reversal), will evolve according to the same physical laws as the present universe. Although this theorem states that the combined CPT symmetry is conserved by the four fundamental forces (gravitation, the electromagnetic force, the strong and the weak force), individual symmetries may be broken. Of interest here is that parity may not be conserved in processes involving the weak force, as suggested by Lee and Yang in 1956[7] and soon after observed by Wu *et al.* in the β-decay of Cobalt-60[8]. The weak interaction is also at play in molecular systems[9-11], notably in the interaction between electrons and nuclei, and the standard model therefore predicts a tiny energy difference between enantiomers of chiral molecules.

The potential energy surface of a chiral molecule shows two minima, which one may associate with the left '*L*' and right '*R*' enantiomers (Figure 1). When the interconversion barrier is very high, the right and left state can be considered to a good approximation as energy eigenstates. In the absence of the weak force those eigenstates are degenerate and the reaction enthalpy $\Delta_r H_0$ for the interconversion reaction between the exact mirror images *L* and *R* enantiomers at 0 K is exactly zero[12]. In the presence of the weak force, a small parity violation energy difference (PVED) $\Delta E_{PV}$ is expected between the ground states (as well as excited states) of the enantiomers (Figure 1). Right- and left-handed molecules cease to be exact mirror images of each other, that is, enantiomers become diastereomers. The PVEDs are, however, predicted to be very small. For the common chiral molecule CHFClBr (**1**, Figure 2), $\Delta E_{PV}$ is predicted to be approximately 30 mHz ($10^{-12}$ cm$^{-1}$ or $10^{-16}$ eV) corresponding to a reaction enthalpy[13] $\Delta_r H_0 \approx 10^{-11}$ J.mol$^{-1}$.

Yamagata[14] then Rein[15], and Gajzago and Marx[16] have first suggested that weak interaction is responsible for an energy difference in the spectrum of right- and left-handed chiral molecules. A number of experimental techniques have been proposed for the observation of PV in molecular systems, including vibrational-rotational[17], electronic[18-20], Mössbauer[21] and NMR[22] spectroscopy, as well as crystallization[23] and solubility[24] experiments, or optical activity measurements[25], but no unambiguous observation has so far been made. The experimental protocol envisaged in this paper, originally proposed by Letokhov[17], consists in measuring differences between two enantiomers, in their infrared spectral absorption line frequencies (Figure 2). As illustrated in Figure 1 and Figure 2, the difference $\Delta \nu_{PV} = \nu_L - \nu_R = (\Delta E_{PV}^* - \Delta E_{PV})/h$ corresponds to a difference of PVEDs between two ro-vibrational levels (*h* is the Planck constant). In 1975 Letokhov suggested that PV would lead to splittings $\Delta \nu_{PV}$ in the vibrational-rotational spectra of enantiomers of chiral molecules on the order of $\Delta \nu_{PV} / \nu \approx 10^{-15} - 10^{-16}$ where $\nu \sim \nu_L \sim \nu_R$ is the transition frequency[17]. His group subsequently searched for such splittings by laser sub-Doppler absorption spectroscopy in the spectrum of racemic CHFClBr[26]. In 1977 Arimondo et al.[27] found that the transition frequencies of the C-C*-CO bending mode of L- and D-camphor (C* denotes a chiral carbon) agreed within their experimental resolution of $10^{-8}$ in relative value. Later on, theoretical studies[28,29] have shown that the PV effect is ten orders of magnitude smaller, that is $\Delta \nu_{PV}/\nu \approx 10^{-19}$.

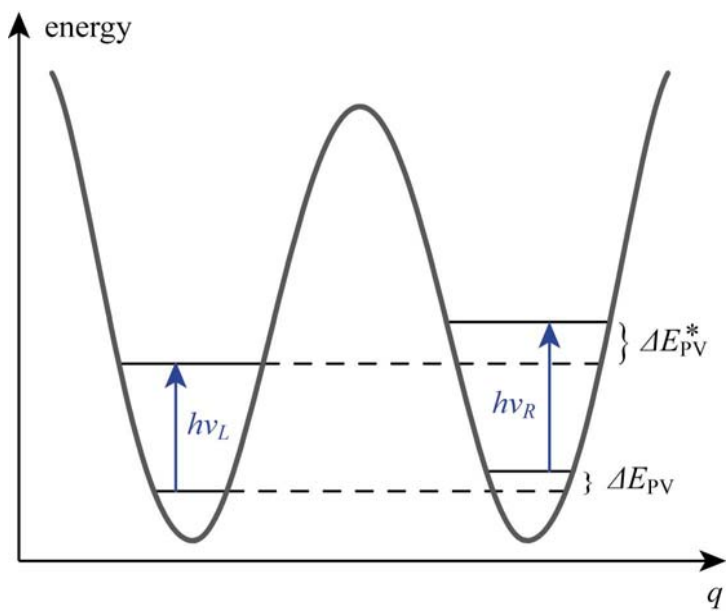

**Figure 1:** The potential energy surface of a chiral molecule shows two minima (as a function of the inversion coordinate $q$), associated with the left 'L' and right 'R' enantiomers. The weak interaction, is responsible for a small parity violation energy difference $\Delta E_{PV}$ between the enantiomers, and consequently a difference between the ro-vibrational frequencies $\nu_L$ and $\nu_R$ of the left and right enantiomers ($h$ is the Planck constant).

In this paper, we review the progress of our joint efforts towards the first experimental observation of PV in molecular systems, to be achieved by high resolution ro-vibrational spectroscopy[30,31], following Letokhov's proposal[17]. This work requires a wide range of competences, from chemistry to physics, and from theory to experiment, and several partners with complementary roles are involved and form a consortium. To observe such a tiny effect as PV, the highest resolution spectroscopy techniques are needed and an experimental set-up based on the powerful method of Doppler-free two-photon Ramsey fringes in a supersonic molecular beam[32] is proposed. Such a device is being set-up by physicists. Meanwhile, the synthesis of enantioenriched molecules that are particularly favorable for PV observation is currently under investigation. Furthermore, theoretical calculations at the relativistic level are being conducted and aim at a better understanding of the origin of PV in molecules as well as guiding the choice of chiral molecules for the experiment. Finally, for a chosen favorable molecule, it is necessary to perform moderately high resolution spectroscopy in order to identify which absorption line will be the best candidate for the PV test. For this purpose, microwave and rovibrational spectroscopy of a candidate molecule are systematically examined.

## II. Basic theoretical considerations

Theory plays a very important role in this field since it can provide predictions of PV effects and thus helps in guiding experimental research. There has been quite a significant activity in the quantum chemical community in the past years with important contributions from the groups of M. Quack, R. Berger, P. Lazzeretti, P. Manninen, P. Schwerdtfeger, L. Visscher and T. Saue that have been summarized in recent reviews[31,33-35].

Conventional quantum chemistry only considers electromagnetic interactions and the molecular Hamiltonian thus commutes with the parity operator

$$[\hat{H}, \hat{P}] = 0$$

which implies that the corresponding molecular wave functions $|\Psi\rangle$ can be characterized by parity. Within the Born-Oppenheimer approximation the solution of the electronic problem provides potential surfaces that are invariant under inversion of the nuclear coordinates. For a chiral molecule two minima of identical well-depth are expected. The electronic wave functions of the enantiomers at the equilibrium structures are related by the parity operation

$$\hat{P}|\Psi_L^{el}\rangle = |\Psi_R^{el}\rangle$$

The solutions to the nuclear problem are states with well-defined parity $|\chi^\pm\rangle$ with equal probability of the equilibrium structures of the two enantiomers. Those states are non-degenerate because of the tunneling splitting. Localized solutions can be generated from linear combinations of the two parity eigenstates $|\chi_{L,R}\rangle = 1/\sqrt{2}(|\chi^+\rangle \pm |\chi^-\rangle)$, but cannot rigourously correspond to eigenstates of the nuclear Hamiltonian since they do not have well-defined parity. However, when the barrier to interconversion (Figure 1) is sufficiently high, enantiomers are kinetically stable and can be considered to a good approximation as degenerate energy eigenstates, as Friedrich Hund pointed out in 1927[36]. While there is no problem concerning this argument, it is less than clear how these kinetically stable left- and right-handed localized molecular states as superpositions between positive and negative parity states are actually prepared (here we can take any time-dependent superposition between the two parity states)[37], and a number of hypotheses has been brought forward in the past[25,38]. Hans Primas expressed this in the following way: "What is the reason that we can buy in the drug store *D*-alanine (state vector $|\Psi_D\rangle$), *L*-alanine (state vector $|\Psi_L\rangle$), but not the coherent superpositions $(|\Psi_D\rangle + |\Psi_L\rangle)$ and $(|\Psi_D\rangle - |\Psi_L\rangle)$?"[39]. Note that for the hypothetical chiral molecule NHDT (as compared to $NH_3$) we would still (to a very good approximation) observe the maser line transition between the two positive and negative parity states. We cannot give an extensive discussion on the evolution of superpositions of chiral states here, which has been intensively discussed and disputed over the last 30 years, but note that in a very recent paper Trost and Hornberger favor the collision hypothesis over other possible mechanisms[40].

In the framework of electroweak chemistry, the picture changes, as two parity violating terms enter the Hamiltonian through the weak interaction between electrons and nuclei. One term depends on the nuclear spin and the observation of its effect on NMR spectra has been considered by several authors[22,41-44]. Here we focus on the observation of parity violation associated with the second term $H_{PV}$. This term is independent of nuclear spin, but induces a minute energy difference between left- and right-handed molecules

$$E_{PV;L} = \langle\Psi_L|\hat{H}_{PV}|\Psi_L\rangle = \langle\hat{P}\Psi_L|\hat{P}\hat{H}_{PV}\hat{P}^{-1}|\hat{P}\Psi_L\rangle = -\langle\Psi_R|\hat{H}_{PV}|\Psi_R\rangle = -E_{PV;R}$$

We obtain a PVED between the two enantiomers $\Delta E_{PV} = 2E_{PV}$. In the framework of quantum field theory all interactions are mediated by virtual bosons. The energy difference between enantiomers originates principally from the exchange of Z bosons between electrons and nuclei. In contrast to photons exchanged in electromagnetic interactions, these bosons are massive, which severely limit the range of the interaction. Parity violation enters the interaction through its dependence on the handedness of the particles, that is the alignment of spin with respect to the direction of motion, and involves almost exclusively left-handed electrons (anti-parallel alignment).

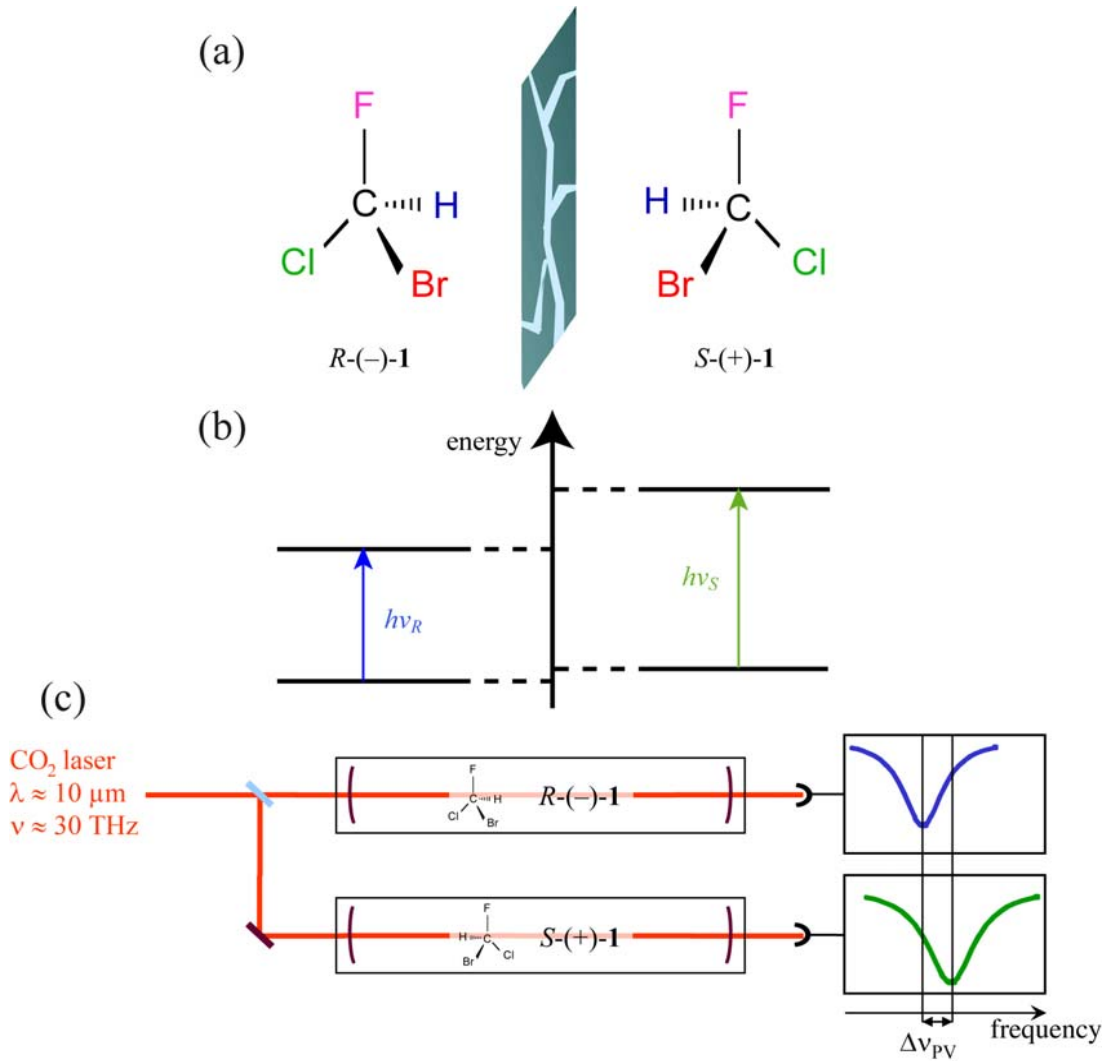

**Figure 2:** Principle of the PV test on CHFClBr. (a) A broken mirror illustrates that owing to PV, CHFClBr enantiomers are not mirror images. (b) This leads to a difference between their ro-vibrational frequencies $\nu_R$ and $\nu_S$. (c) Spectra of the two enantiomers are recorded simultanously using identical cavities. Note that we expect $\Delta\nu_{PV}$ to be much smaller than the line width.

The parity-violating energy can be written as a sum of atomic contributions

$$E_{PV} = \sum_A E_{PV}^A; \quad E_{PV}^A = \frac{G_F}{2\sqrt{2}} Q_{W;A} \langle \Psi | \gamma_5 \rho^A | \Psi \rangle; \quad \gamma_5 = \begin{bmatrix} 0_2 & I_2 \\ I_2 & 0_2 \end{bmatrix}$$

in which appears the weak charge $Q_{W;A} = -N_A + Z_A(1-4\sin^2\theta_W)$, where $Z_A$ and $N_A$ are the number of protons and neutrons, respectively, of nucleus A and where $\theta_W$ is the Weinberg mixing angle (the most recent value is $\sin^2\theta_W = 0.2397(13)$ [45]). The appearance of the normalized nuclear density $\rho^A$ implies that the integration over electron coordinates is restricted to nuclear regions. The same holds true for the $\gamma_5$ matrix since it couples large and small components of Dirac 4-spinors, due to the highly atomic nature of the small components. The Fermi coupling constant $G_F = 2.22254 \times 10^{-14} E_h a_0^3$ demonstrates the minuteness of the PV effect ($E_h$ is Hartree energy and $a_0$ is the Bohr radius). The parity-violation energy difference is strictly zero for atoms and achiral molecules, and for chiral molecules in the absence of spin-orbit coupling. On the other hand, the energy scales approximately as $Z_A^5$, which has directed interest towards molecules containing heavy atoms[46,47]. Thus, parity violation in molecular systems will be better treated in a relativistic framework.

## III. PV experiments on CHFClBr

The first high-sensitivity test of parity violation in molecules was performed with CHFClBr (**1**), the molecule suggested by Letokhov *et al.*[26]. It was first synthesized in high optical purity in 1989 by Doyle and Vogl[48]. Its *R*-(−)/*S*-(+) absolute configuration was determined in 1997 by A. Collet, J. Crassous and co-workers[49,50]. Following this work, a first high resolution experiment on *S*-(+)-**1** and *R*-(−)-**1** with respective enantiomeric excesses of 56.5% and 22%, was carried out in 1999 by the group of C. Chardonnet using laser-saturated absorption spectroscopy in two Fabry-Perot cavities (Figure 2). A $CO_2$ laser-based spectrometer was developed to probe a hyperfine component of the C-F stretching fundamental band of $CHF^{37}Cl^{81}Br$ at ~30 THz (~10 µm or 1000 cm$^{-1}$). The spectrum of each enantiomer was simultaneously recorded in separate Fabry-Perot cavities. Thus, the sensitivity of the experiment was governed by the precision of determination of the line center and not by the spectral line width, which was not better than 60 kHz. Over 10 days of measurement a mean difference $\Delta\nu_{PV}^{R(-)/S(+)} = \nu_{R(-)} - \nu_{S(+)}$ of 9.4 Hz was obtained, with statistical and systematic uncertainties of 5.1 Hz and 12.7 Hz respectively[51]. This gives an upper bound of $\Delta\nu/\nu \approx 4\times10^{-13}$ for the parity violation effect. These experiments were repeated in 2002 with samples of higher enantiomeric excess (*S*-(+)-**1**: 72% and *R*-(−)-**1**: 56%) and a stronger and narrower CHFClBr hyperfine transition[52]. Over 7 days of measurement an experimental sensitivity of 5×10$^{-14}$ was obtained. These latter measurements revealed that the mean difference was proportional to pressure, which is not expected for a PV effect. This was attributed to uncontrollable residual gases present in the absorption cells (at a level below 5%) and responsible for uncompensated systematic effects due to collisional shifts of the transition frequency. Furthermore, even though the experimental sensitivity was improved by five orders of magnitude compared to the previous experiment on camphor[27], later theoretical studies predicted that the PV difference for the C-F stretch of CHFClBr is on the order of -2.4 mHz[53,54], corresponding to $\Delta\nu_{PV}^{R(-)/S(+)}/\nu \approx -8\times10^{-17}$. Therefore, experimental detection of a PV shift for compound **1** would need an improvement of three orders of magnitude in sensitivity. Thorough analysis of the experimental results showed the limits of a set-up based on saturated absorption spectroscopy of the enantiomers filling at low pressure (a few tenths of Pa) two identical Fabry-Perot cavities: (i) the too large width (60 kHz) of the probed line, (ii) the low enantiomeric purity of the samples and (iii) above all the presence of residual impurities in the samples, responsible for a systematic effect via collisional shifts of the transition frequencies. It is practically impossible to reduce this collisional effect by more than one order of magnitude, which would not improve enough the sensitivity, for a PV test on most of the molecules of interest. In order to go further, it is necessary to develop a new experiment in which collisional effects are negligible.

## IV. A new project

In this context, supersonic molecular beam spectroscopy using the powerful ultra-high resolution technique of Doppler-free two-photon Ramsey fringes, as recently developed in C. Chardonnet's team on $SF_6$ for molecular frequency metrology, seems very promising[55,56]. The Laboratoire de Physique des Lasers (LPL) group thus decided to develop such an experimental set-up adapted to the spectroscopy of chiral molecules for a new PV experiment. The principle is similar to the previous experiment on CHFClBr: comparing the frequency of the same ro-vibrational line for two enantiomers of a chosen chiral molecule. Any frequency difference is interpreted as a PV effect which can only be due to the weak interaction.

Supersonic beam molecular spectroscopy was introduced by W. A. Klemperer[57] and presents the major advantage to get rid of intermolecular collisions, which is responsible for both a frequency shift and a broadening of the molecular line. In a supersonic beam there is virtually no collision inside the beam. Thus, collisions are mainly due to background pressure which is 4 to 5 orders of magnitude lower than in the cell experiment performed on CHFClBr.

A supersonic beam is much more interesting than a regular thermal beam since, due to the many collisions occurring at the very beginning of the supersonic expansion, the rotational temperature drops down to a few K (thermal and internal energy are converted into a large common translational kinetic energy). As a consequence lower energy levels are strongly populated which turns out to amplify the spectroscopic signal when those levels are probed.

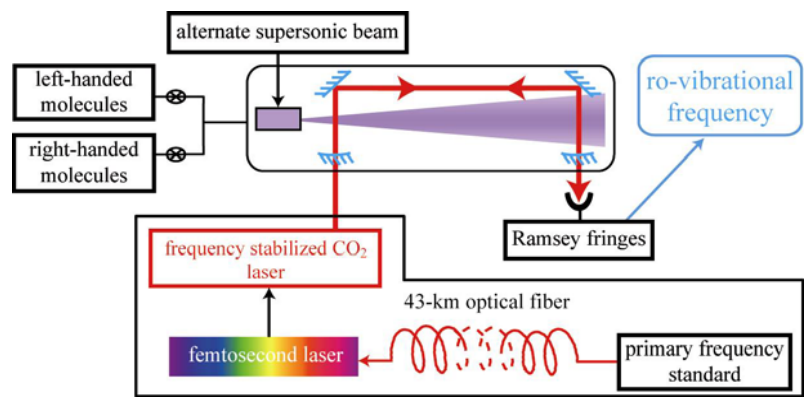

**Figure 3:** Principle of the Ramsey fringes experiment considered for the observation of PV in molecules. A frequency-controlled $CO_2$ laser interrogates an alternate beam of right- and left-handed molecule.

For observing a PV effect, we aim to reach the narrowest possible line, since, for most systematic effects, the uncertainty on the line frequency measurement is proportional to its width. We plan to use two-photon spectroscopy which, just like saturated absorption spectroscopy is a way to avoid Doppler broadening, that strongly limits the experimental resolution. For this purpose, molecules will be probed by two counter-propagating beams where they will absorb two identical photons traveling in two opposite directions. The corresponding Doppler frequency differences cancel out and the resonance condition is satisfied whatever the molecule velocity is: this is called Doppler-free two-photon spectroscopy. Note that the two-photon transition will connect $v = 0$ and $v = 2$ vibrational levels of the chiral molecule. In a molecular beam experiment, the line width is then limited by the inverse of the transit time through the ~1 cm-diameter laser beam, that is typically 100 kHz, for a molecular velocity of about 1000 m/s. To narrow the line even more and thus improve the resolution, we use the method of separated fields introduced by Ramsey[58]. Combining two-photon spectroscopy and the Ramsey technique requires only two phase-related standing waves[59,60], which is quite easy to implement with a folded Fabry Perot cavity, as shown on Figure 3. In particular, no strict parallelism between the standing waves is required. Molecules will successively cross the two laser interrogation zones which will lead to a fringe pattern in the absorption signal, when scanning the laser detuning around the resonance. The fringes periodicity corresponds to half the inverse of the propagation time through such a molecular interferometer. A typical 1 m distance between interrogation zones gives a periodicity of a few 100 Hz, leading to two to three orders of magnitude improvement in the resolution compared to a single laser beam set-up and compared to the cell experiment performed on CHFClBr (**1**, see section 0).

As shown on Figure 3, for the LPL experiment dedicated to the PV observation, we propose to use the same set-up, alternatively fed with right- and left-handed molecules. The spectra of both enantiomers will then be recorded in identical experimental conditions (pressure, temperature,...), with the same set-up (nozzle, skimmer,...). We expect the interaction of the right- and left-handed molecules with the laser zones to be identical, such a differential measurement thereby assuring a high degree of cancellation of systematic errors.

Compared to the experiment on CHFClBr, the left- and right-handed molecules will not be probed simultaneously but by turns, with the risk to face a laser frequency drift during the experiment. This is the price to be paid for almost perfectly identical interaction geometry and conditions for the two enantiomers. However, the LPL group will benefit from the tremendous progress made in the recent years for the laser frequency control. The reference laser is actually connected to the primary frequency standard located at the laboratory LNE-SYRTE (Laboratoire National d'Essais-Systèmes de Référence Temps Espace) at Paris Observatory, consisting of a cesium fountain and a hydrogen maser, whose frequency accuracy reaches nowadays a few parts in $10^{-16}$ in fractional value. A metrological reference frequency generated at LNE-SYRTE arrives at LPL via a 43-km long optical fiber, where it is compared to the $CO_2$ laser by means of a frequency comb generated by a mode-locked femtosecond Ti:Sa laser (Figure 3). Recent results show that the residual noise added by the link is at the level of $7\times10^{-20}$ over a few hours and $2\times10^{-18}$ over 100 seconds[61,62]. Finally, the LPL $CO_2$ laser frequency control is limited by laser frequency noise to about 0.1 Hz after a few minutes and shows an accuracy and stability over a day of $10^{-16}$ (~10 mHz), limited by that of the LNE-SYRTE reference frequency[55,56,63]. As a conclusion, we do not consider that this will limit our molecular PV experiment.

Figure 4 shows an example of a recording of the central fringes obtained at LPL on the Ramsey fringes experiment mentioned above using a beam of $SF_6$. It enabled us to measure the sulfur hexafluoride absolute frequency with a sub-Hz ($10^{-14}$) uncertainty[55,56]. The ultimate accuracy of the differential measurement we plan for the PV test is, of course, a key question. We must here distinguish between systematic and statistical uncertainties. A realistic estimate of the systematic uncertainties is difficult to do but at least, at this stage, we can compare with both the previous experiment on CHFClBr and the beam experiment on $SF_6$. First, the background pressure which limited the previous experiment to a few Hz is not an issue in a molecular beam experiment. Second, the probed line width will be of the order of 100 Hz instead of 60 kHz, which reduces a number of systematic errors accordingly. Finally, a number of systematic effects will be identical for left- and right-handed molecules and will thus cancel out after our differential measurement. Reducing the systematic uncertainties below 0.01 Hz ($3\times10^{-16}$) seems thus realistic with this scheme. The main limitation is expected to be the statistical uncertainty which usually decreases as the square root of the integration time (for a white noise), but is proportional to the inverse of the signal-to-noise ratio that we cannot infer yet. For that reason, spectroscopic studies and theoretical investigations of promising molecules will be of great help to identify candidates for a PV observation. We will first focus attention on molecules which present resonances within the $CO_2$ laser spectral window although this constraint could be lifted later on if it turns out to be too strong.

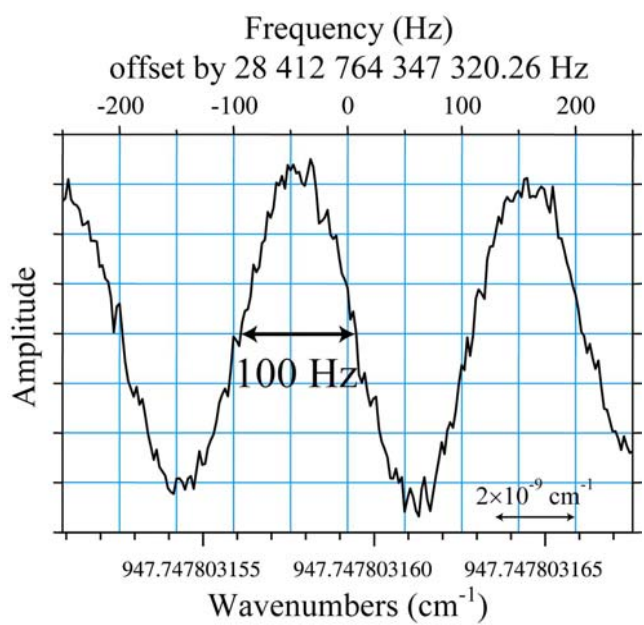

**Figure 4:** Central Ramsey fringes obtained with a pure $SF_6$ supersonic beam. The resolution is 100 Hz and the signal to noise ratio is 30 averaging 1 s per point (200 points) [55,56]. The SF6 absolute ro-vibrational frequency was measured to be
28 412 764 347 320.26 ± 0.79 Hz (corresponding to 0 Hz on the frequency scale).

## V. Theoretical studies

A computational protocol for the rapid screening of candidate molecules has been developed by R. Bast and T. Saue[64], based on 2-component relativistic density functional theory (DFT) calculations:

1. *Simulation of infrared spectra :* A geometry optimization followed by a rovibrational analysis produces a list of harmonic frequencies and corresponding intensities as well as rotational constants. The frequencies and the intensities are used to simulate the infrared spectrum of the candidate molecule (see for example complex **10** in Figure 5) and to identify the vibrational modes in the operating range of the $CO_2$ laser. The calculations furthermore allow the identification of low-frequency modes and small rotational constants that would degrade the resolution of the experiment. If a molecule appears promising, the potential curve is calculated along the selected vibrational mode and the vibrational problem is solved. Anharmonicities, but not intermode couplings[53], are thus included in our theoretical model. Calibration studies show that the electronic structure calculations in this step can be carried out at the DFT (B3LYP) level including relativistic effects through pseudopotentials.

2. *Calculation of the PV contribution to infrared spectra:* In this step the energy $E_{PV}(q)$ is calculated along the selected normal mode coordinate $q$ and then integrated with the vibrational wave functions to obtain the PV frequency difference for the lower vibrational levels of the mode. Our calibration studies show that the exact Two-Component Hamiltonian (X2C), developed by M. Iliaš and T. Saue[65], coupled with two-electron spin-orbit corrections by the atomic mean-field integrals (AMFI) approach[66,67] is adequate for this step.

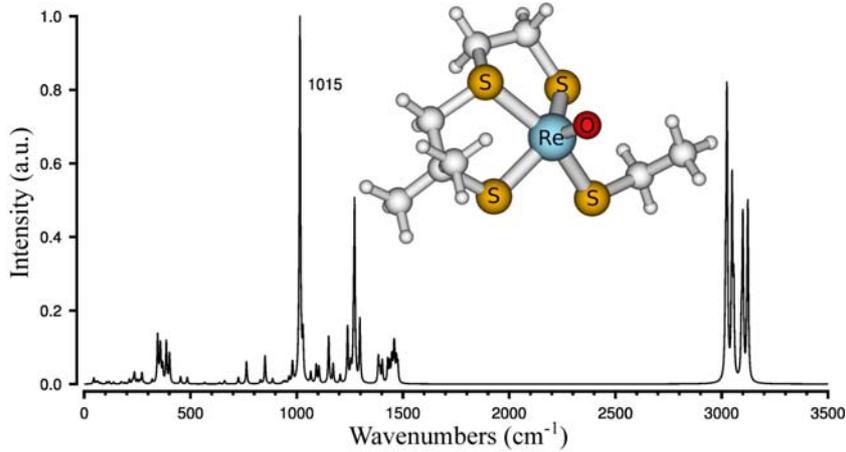

**Figure 5:** Simulated infrared spectrum of rhenium complex **10**.

Parallel to the screening of candidate molecules we are carrying out theoretical studies aiming at a deeper understanding of parity violation in chiral molecules. Several approaches are being pursued. One line of attack is to consider the PV frequency shift $\Delta\nu_{PV}$ in the framework of perturbation theory. From second-order perturbation theory, starting from harmonic solutions, the PV shift in the fundamental transition is given by[54]:

$$\Delta\nu_{PV} \approx \frac{\hbar}{\mu\omega_e}\left[P^{[2]} - \frac{1}{\mu\omega_e^2}P^{[1]}V^{[3]}\right]$$

where $V^{[n]}$ and $P^{[n]}$ are the MacLaurin expansion coefficients of the potential and $E_{PV}(q)$ along the selected normal coordinate $q$, respectively, $\mu$ is the reduced mass and $\omega_e$ is the vibrational frequency. It is evident from this expression that the essential quantity is not the electronic value of the PV energy at equilibrium $P^{[0]}$, rather its variation along the vibrational coordinate. For instance, one immediately sees that a linear variation of the PV energy ($P^{[2]} = 0$) combined with a strictly harmonic mode ($V^{[3]} = 0$) leads to zero PV frequency difference. Furthermore, since the cubic force constant $V^{[3]}$ is generally negative, the harmonic and anharmonic contributions to the PV vibrational frequency difference tend to cancel if $P^{[1]}$ and $P^{[2]}$ have opposite signs. While there is some understanding of factors affecting the size of $P^{[0]}$ in molecular system (see for instance Refs. 47 and 68), such knowledge is mostly unchartered territory for the derivatives $P^{[n]}$ ($n > 0$) with respect to the selected normal coordinate. Another line of attack is based on the decomposition of the expectation value $E_{\text{PV}}$ into intra- and interatomic contributions. A preliminary result from such analysis is that the contribution $E_{\text{PV}}^{\text{A}}$ to the PVED associated with a given nucleus $A$ is completely dominated by intraatomic contributions from the same center, more precisely from the mixing of (valence) $s_{1/2}$ and $p_{1/2}$ orbitals by the field of the surrounding atoms. This finding strongly suggests that it should be possible to develop a LEGO model of parity violation in molecular systems. The idea is to combine pre-calculated atomic quantities with simple bonding models such that the PV energy and, most important, its variation with molecular geometry can be predicted. If successful, such a model would allow a rapid screening of candidate molecules and possibly computer-aided design of new molecules optimized for the high resolution experiment.

## VI.  Molecules considered for a new PV test by the consortium: a brief review

From the $Z_{\text{A}}^5$ scaling law for the PVED evoked in section II, it is clear that candidate molecules should preferably contain one or more heavy atoms at or near the stereochemical center. Accurate relativistic *ab initio* calculations were performed on the fluorohalogenomethanes CHFClBr (**1**), CHFClI (**2**), CHFBrI (**3**) and CFBrClI (**4**) by P. Schwerdtfeger and coworkers[54,69,70]. A set of parity-violating differences in frequencies between *R* and *S* enantiomers were obtained for the C-F stretching fundamental transition. They are shown in Table I. It turns out that the PV effect in the C-F stretching band of CHFClBr (**1**) is accidentally small and is dominated by the anharmonic part of the vibrational motion[69]. CHFBrI seems to be the most promising molecule of this series with $\Delta \nu_{\text{PV}}^{R/S} = -50.8$ mHz. Recently Faglioni and Lazzeretti considered BiHFBr and BiHFI[71]. The PV frequency shift $\Delta \nu_{\text{PV}}$ could possibly reach 20 Hz, but these species are thermodynamically unstable and one needs to consider the first overtone of the H–Bi–X bending mode in order to reach the frequency range (~30 THz) of a $CO_2$ laser spectrometer. Lately, chiral selenium, tungsten, gold, mercury, iridium, osmium and rhenium complexes have been calculated to be favorable candidates for PV observation by P. Schwerdtfeger, R. Bast and coworkers[72-77]. For instance, the two compounds, $Os(\eta^5\text{-Cp})(=CCl_2)Cl(PH_3)$ (**5**) and $Re(\eta^5\text{-Cp*})(=O)(CH_3)Cl$ (**6**) (Figure 6), both present a strong vibrational band (respectively Os=C and Re=O stretching mode) in the 30 THz region of the $CO_2$ laser, and a large PV effect $\Delta \nu_{\text{PV}}$ around 1 Hz[74] (obtained from 4-component relativistic Hartree-Fock calculations). This theoretical result is very promising for our project. In addition, for synthetic chemists, it is conceivable to produce similar chiral-at-metal complexes in large quantity and with an enantiomeric excess of 100%. One should note however that these molecules are solid at room temperature, which constitutes a difficulty for high resolution gas phase spectroscopy.

**Table I:** Frequency difference $\Delta \nu_{PV}^{R/S} = \nu_R - \nu_S$ of the C-F stretching vibrational mode of a series of fluorohalogenomethanes[54,70] from 4-component relativistic Hartree-Fock (HF) calculations. At the MP2 level the frequency difference $\Delta \nu_{PV}^{R/S}$ for CHFClBr is modified to -1.7 mHz[70].

| CHFClBr (**1**) | CHFClI (**2**) | CHFBrI (**3**) | CFClBrI (**4**) |
|---|---|---|---|
| -2.4 mHz | -23.7 mHz | -50.8 mHz | +11.6 mHz |

The ideal candidate chiral molecule for the experiment described in section IV should satisfy the following requirements:
- show a large PV vibrational frequency difference $\Delta \nu_{PV}$ of an intense fundamental band preferably within the $CO_2$ laser operating range (850-1120 cm$^{-1}$);
- be available in large enantiomeric excess or, ideally, in enantiopure form;
- not be too bulky since the sensitivity of the experiment will be largely determined by the partition function of the molecules (note that in a supersonic beam, the internal degrees of freedom are frozen down to about 1 K);
- avoid nuclei with a quadrupole moment in order to avoid large hyperfine structure;
- have a suitable two-photon transition joining a state in the fundamental vibrational level, $\nu = 0$ to one in the $\nu = 2$ level;
- allow the production of the supersonic expansion, (thus sublimate without decomposition for solid state molecules; laser ablation techniques may also be envisaged);
- be available at gram-scale.

In the light of those requirements, our consortium first took up with the synthesis and the study of fluorohalogenomethanes heavier than CHFClBr, especially chlorofluoroiodomethane. Later on, since it had been shown that organometallic complexes containing heavier atoms may display much higher PV effects, we focused on the synthesis of chiral oxorhenium compounds.

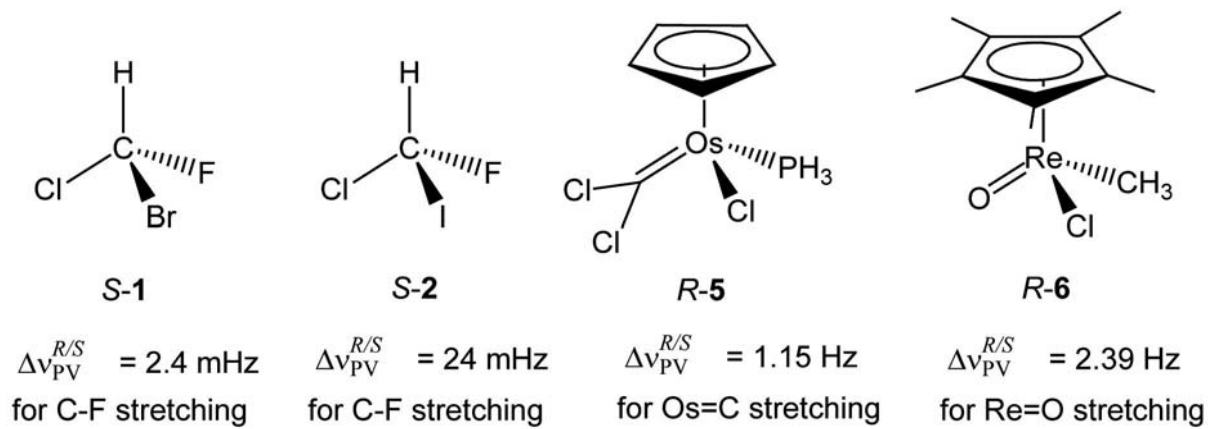

**Figure 6:** Comparison of PV effects $|\Delta \nu_{PV}|$ for halogenomethanes (**1**, **2**) and transition metal complexes (**5**, **6**), obtained from 4-component relativistic HF[54,70] and density functional theory (DFT)[64] calculations, respectively.

## A. *Chlorofluoroiodomethane (CHFClI)*

Chlorofluoroiodomethane has been considered for a while as a potential candidate, because it has several interesting features which could make a new experimental PV test possible on this molecule[30,31]. First, its calculated PV difference is one order of magnitude larger than for CHFClBr. Second, the high vapor pressure of CHFClI allows investigation by supersonic beam spectroscopy. Finally, as detailed below, ground-state rotational constants have been determined rather accurately by microwave and millimeter wave spectroscopy[78,79]. Even if excited state constants have been estimated by infrared spectroscopy with less accuracy[79], this set of spectroscopic data significantly helps the selection of the most appropriate absorption line for a PV test. Due to the low stability of iodinated compounds, the other halogenomethanes CHFBrI and CFClBrI have not been prepared yet.

### a) Synthesis and resolution of CHFClI

Racemic CHFClI (**2**) was synthesized in multigram quantities for spectroscopic studies. Then partially resolved (+) and (–)-CHFClI enantiomers with respectively 63% and 20% enantiomeric excesses (*ee*) were prepared by decarboxylation of partially resolved diastereomerically enriched (–)-strychnine salts of chlorofluoroiodoacetic acid (Scheme 1). Diastereomerically enriched *n* and *p* salts {(+)-IFClCCO$_2$H,(–)-strychnine} and {(–)-IFClCCO$_2$H,(–)-strychnine} were obtained by simple crystallization in methanol. Their decarboxylation to (+) and (–)-**2** was conducted in a protic solvent (triethyleneglycol, TEG). Direct distillation under reduced pressure enabled to obtain **2** as a pure liquid. The *ee* values were measured by molecular recognition using chiral hosts (cryptophanes)[80] or by analytical gas chromatography on a Chirasil-γ-Dex column under cryogenic conditions[81]. The *S*-(+)/*R*-(–) absolute configuration was ascertained by vibrational circular dichroism (VCD) in the gas phase[79].

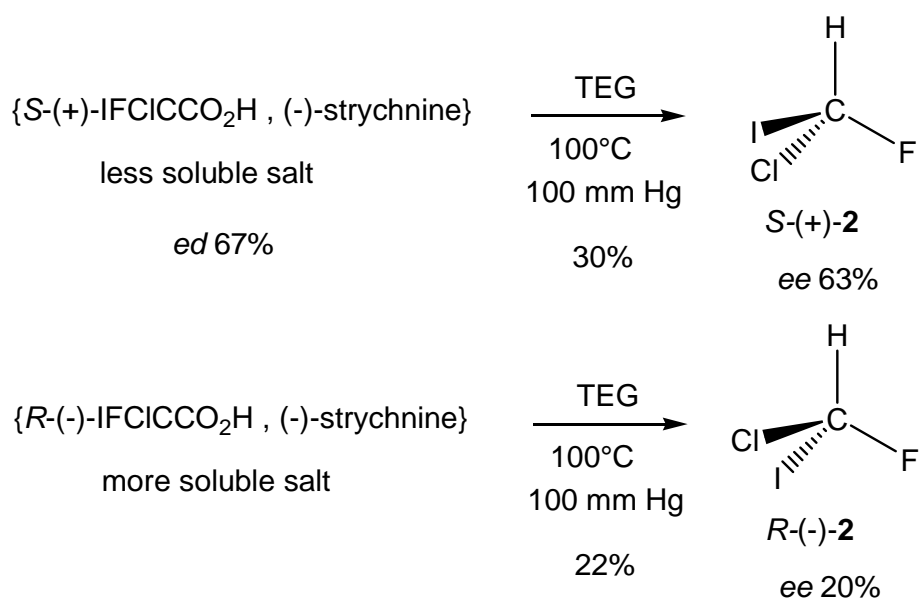

**Scheme 1:** Synthesis of *S*-(+) and *R*-(–) chlorofluoroiodomethane (**2**).

### b) Spectroscopic study of CHFClI

Microwave and infrared spectroscopies have been performed in order to analyze precisely the rovibrational band and to enable the selection of the most appropriate absorption line for a PV test. The rotational spectrum of chlorofluoroiodomethane (CHClFI) has been investigated in T. Huet's group[78,79]. Because its rotational spectrum is extremely crowded, extensive *ab initio* calculations were first performed in order to predict the molecular parameters. The low $J$ transitions ($J$ is the rotational quantum number) were measured using a pulsed-molecular-beam Fourier transform spectrometer, and the millimeter-wave spectrum was recorded to determine accurate centrifugal distortion constants. Because of the high resolution of the experimental techniques, the analysis yielded accurate rotational constants, centrifugal distortion corrections, and the complete quadrupole coupling tensors for the iodine and chlorine nuclei, as well as the contribution of iodine to the spin-rotation interaction. These molecular parameters were determined for the two isotopologs $CH^{35}ClFI$ and $CH^{37}ClFI$. They reproduce the observed transitions within the experimental accuracy. Moreover, the *ab initio* calculations have provided a precise equilibrium molecular structure, and the *ab initio* molecular parameters are found in good agreement with the corresponding experimental values.

Rovibrational Fourier Transform InfraRed (FTIR) spectroscopy of the $\nu_4$ C-F stretching band has been performed in the group of P. Asselin. Molecular jet experiments as well as spectroscopy in a static cell at room temperature have been performed on the fundamental $\nu_4$ and the first overtone $2\nu_4$ band of CHFClI. The principle of the infrared jet-cooled experiment has been described elsewhere[79]. Figure 7 displays a characteristic *PQR* rovibrational structure of the $\nu_4$ band of **2** recorded at the same resolution (0.008 cm$^{-1}$) in the cell (a) and in a supersonic expansion (b). The stick spectrum (Figure 7 (c)) displays possible coincidences with $P(J)$ and $R(J)$ lines of the 9.4 μm band of the CO$_2$ laser. Due to the too small available quantity of **2** the quality and the resolution of the infrared jet spectrum were not high enough to perform a precise rovibrational analysis of the $\nu_4$ band.

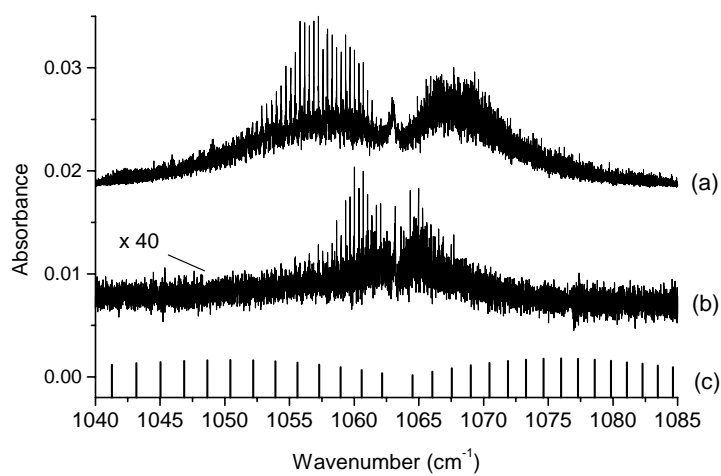

**Figure 7:** FTIR spectra of the $\nu_4$ band of CHFClI (**2**) recorded at 0.008 cm$^{-1}$ resolution (a) in a room temperature static cell (b) in a supersonic expansion. (c) Stick spectrum of $P(J)$ and $R(J)$ lines of the 9.4 μm band of the CO$_2$ laser.

The cell spectrum of the first overtone $2\nu_4$ band provides two useful pieces of information: the intensity ratio between $0 \to \nu_4$ and $0 \to 2\nu_4$ transitions, estimated to about 200, and a measure of the deviation from the resonance for two-photon transitions, i.e. $2\nu_4 - 2(\nu_4)$, as large as 17 cm$^{-1}$. This unfortunately rules out the possibility of finding an appropriate two-photon transition for ultra-high resolution experiments based on the Doppler-free two-photon Ramsey fringes technique (see section IV).

These experiments demonstrated the feasibility of high resolution supersonic beam spectroscopy of molecules in a liquid phase at ambient temperature. However, in addition to the impossibility of finding an appropriate two-photon transition, the molecule CHFClI (**2**) is unfortunately not stable enough and to date, synthesis in enantiopure form and gram quantity can hardly be achieved. For these reasons, we decided to switch to chiral oxorhenium complexes, mentioned in section 0 for their high PV effect and for the feasibility to produce such molecules in large quantity with an enantiomeric excess of 100%.

## B. *Chiral oxorhenium complexes*

The transition metal complexes **5** and **6** discussed in section 0 and shown in Figure 6 have limited utility for the proposed PV experiment since they are expected to easily racemize due to the presence of the pentamethylcyclopentadienyl ligand (Cp*). Two more promising families of chiral oxorhenium complexes, based on either hydrotris(1-pyrazolyl)borate (Tp) or sulfur ligands have thus been synthesized by J. Crassous, L. Guy and coworkers, and characterized by VCD as well as anomalous X-ray diffraction. They present an intense band around 1000 cm$^{-1}$, associated with the Re=O stretching mode.

### a) **Tp ligand-based complexes**

The first studied oxorhenium complexes were the two enantiomeric pairs $S_{Re},S_{O-C},R_{C-N}$- and $R_{Re},R_{O-C},S_{C-N}$-[TpReO($\eta^2$-N(CH$_3$)CH(CH$_3$)CH(Ph)O-*N,O*)] **8**, because they were readily synthesized according to literature procedures from the precursor TpReOCl$_2$ **7** and ephedrine as the chiral ligand[82,83]. Due to steric interactions, the reaction was stereoselective since only one diastereomer was obtained among the two possible ones. In other words, the chirality was transferred from the ephedrine ligand to the stereogenic rhenium center. The absolute configuration was determined by Faller and Lavoie by X-ray crystallography[82] and was subsequently confirmed by our VCD spectroscopy studies. Indeed, as shown in Scheme 2, the strong infrared band at 930 cm$^{-1}$ corresponding to the Re=O stretching mode is VCD active.

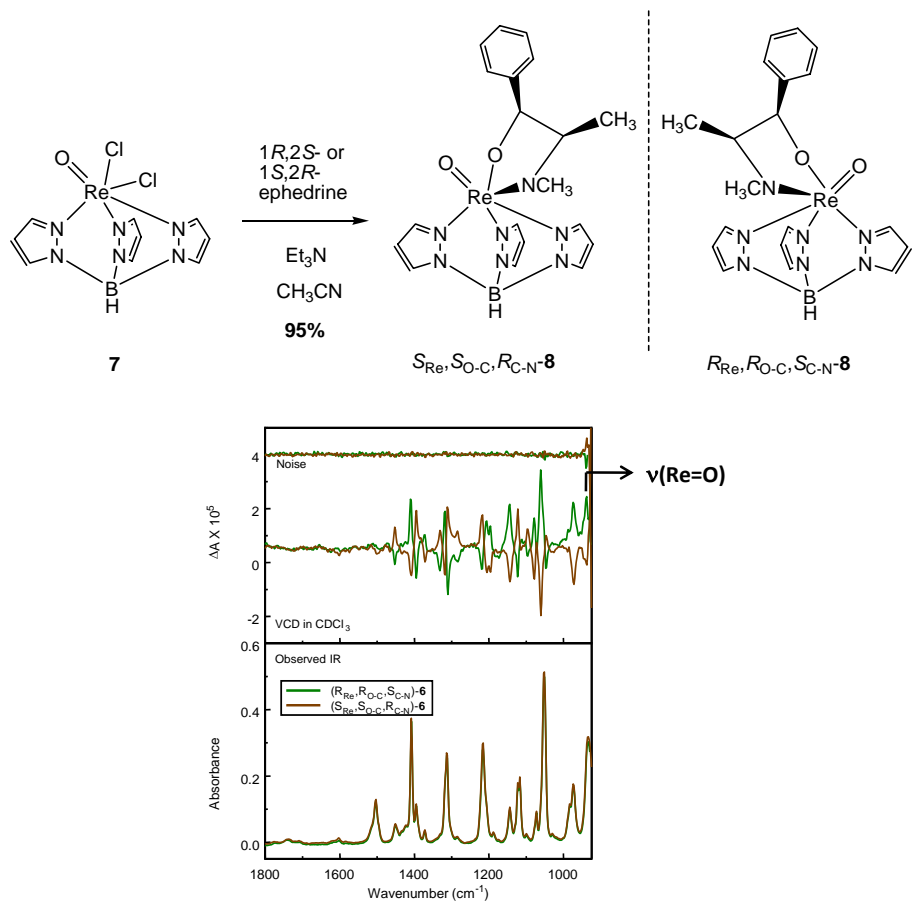

**Scheme 2:** Synthesis of Tp-based chiral oxorhenium enantiomeric complexes **8** and their infrared and VCD spectra.

Tp ligand-based complexes present the advantage of having an intense band around 1000 cm$^{-1}$, of being synthesizable at the scale of several grams in enantioenriched form and can be sublimated. Nevertheless, they are too bulky which makes the spectroscopic analysis and the calculation of the PV effect complicated.

### b) Sulfur ligand-based complexes

#### (i) Synthesis and resolution of enantiomers

In order to simplify the chemical structure of our candidate molecule for PV tests, we then focused on the preparation of complexes **10-16** (Scheme 3-Scheme 5 and Figure 8), which belong to the family of « 3+1 » mixed complexes, bearing a tridentate ligand (1,1-dimethyl-1,5-dithiol-3-thiapentane) and a monodentate one.

#### (i) Synthesis and resolution of enantiomers

The synthesis is based on the use of oxorhenium precursor, ReOCl$_3$(PPh$_3$)$_2$ (**9**), which reacts with the tridentate and monodentate ligands in the presence of sodium acetate in THF. This versatile method allows the preparation of a large panel of complexes bearing thioalkyle (**10-11**), iodo (**12**), SPh-4-Br (**13**), SPh (**14**) and SePh (**16**) groups as described in Scheme 3-Scheme 5[64,84]. In all complexes, the rhenium atom is a stereogenic center since the tridentate ligand is dissymmetric due to the presence of the gem-dimethyl. Note that in these complexes, the chirality only comes from the presence of the two methyl groups. Interestingly sulfidorhenium complex **15** only bears Re, C, H and S atoms and is chiral. It was prepared from a sulfuration reaction of oxorhenium **14** with $P_4S_{10}$. The structure and absolute configuration of all complexes was ascertained by anomalous X-ray crystallography. Due to their neutrality, they were readily resolved in their enantiopure form by using HPLC separations over chiral stationary phases. It was therefore possible to obtain several hundreds of milligrams of pure enantiomers.

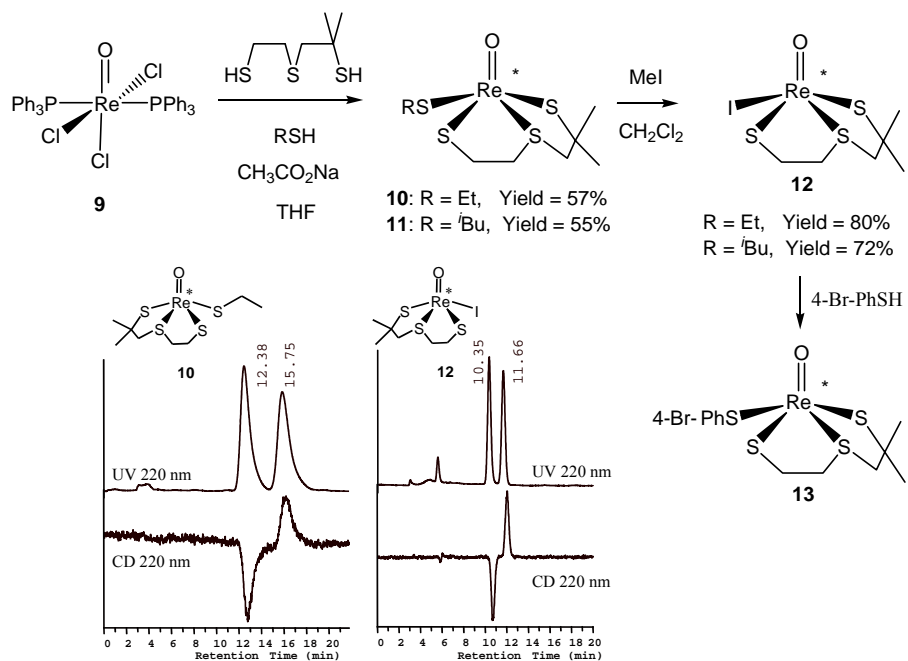

**Scheme 3:** Synthesis of mixed « 3+1 » oxorhenium complexes **10-13** and their resolution by HPLC over a chiral stationary phase. Chiralpak AS-H, Hexane/ethanol (1/1) – 1 ml/min, Detection : UV and CD at 220 nm.

Enantimerically enriched samples of iodo complex **12** (*ee* 89%) were obtained directly from enantiomerically pure **10** (Scheme 4) by reaction with methyliodide in dichloromethane. This reaction occurred with retention of the configuration at the rhenium atom. Such a reaction enables direct access to a variety of enantioenriched mixed « 3+1 » oxorhenium complexes such as **13** in Scheme 3 by replacing iodine with other nucleophiles.

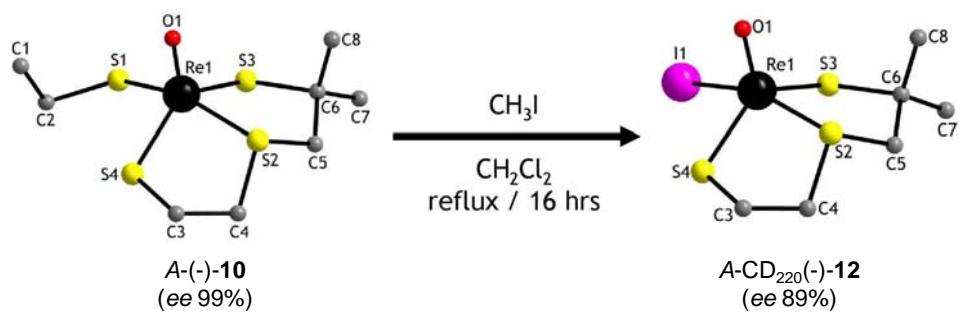

**Scheme 4:** Stereoselective synthesis of enantioenriched chiral oxorhenium complex *A*-**12** from *A*-**10** (for the stereochemical descriptors see Refs. 85 and 86).

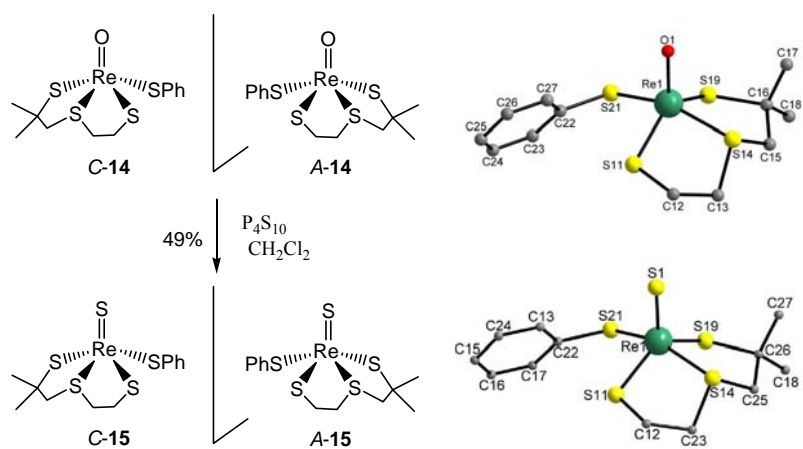

**Scheme 5:** « 3+1 » mixed oxo- and sulfido-rhenium complexes **14**-**15** based on the aromatic SPh ligand.

VCD spectroscopy studies of complex **16** give more insight into the chiral environment of the metal center. Interestingly we have observed that the rotational strength of the Re=O and Re=S stretching bands around 950 cm$^{-1}$ and 510 cm$^{-1}$ respectively were significantly influenced by the chiral conformation adopted. Indeed, conformational analysis of compound **16** revealed two conformers **C1** and **C2** of similar stability, with O-Re-Se-C dihedral angles +60° and −60° as depicted in Figure 8. The most intense infrared and VCD band at 956 cm$^{-1}$ corresponds to the Re=O stretching mode and an anisotropy ratio for conformer **C2** of moderate size, $4R/D = \Delta\varepsilon/\varepsilon = +7.3 \times 10^{-5}$. This band exhibits VCD intensity with opposite sign and nearly the same magnitude for conformers **C1** and **C2**. These VCD intensities are indicative of the near mirror environments about the rhenium arising from the orientations of the Se-phenyl group, and reflect the minor effect of the methyl groups on this vibration. The overall agreement between observed and calculated spectra is excellent, giving the absolute configuration *C*-ECD$_{220}$(+)-**16** and *A*-ECD$_{220}$(-)-**16**.

From this study it becomes obvious that VCD spectroscopy is a precious tool to carefully examine the chiral environnement around the rhenium atom and its influence on the Re=O stretching mode. A highly chiral environement at the metal center is a key-point for the future parity violation observation since it will induce large parity violation energy differences. As shown from theoretical calculations, the two methyl groups are not sufficient to induce a high dissymmetry on complexes **10**-**16**, leading to $\Delta E_{PV}$ values one order of magnitude smaller than the original model oxorhenium complex **4**.

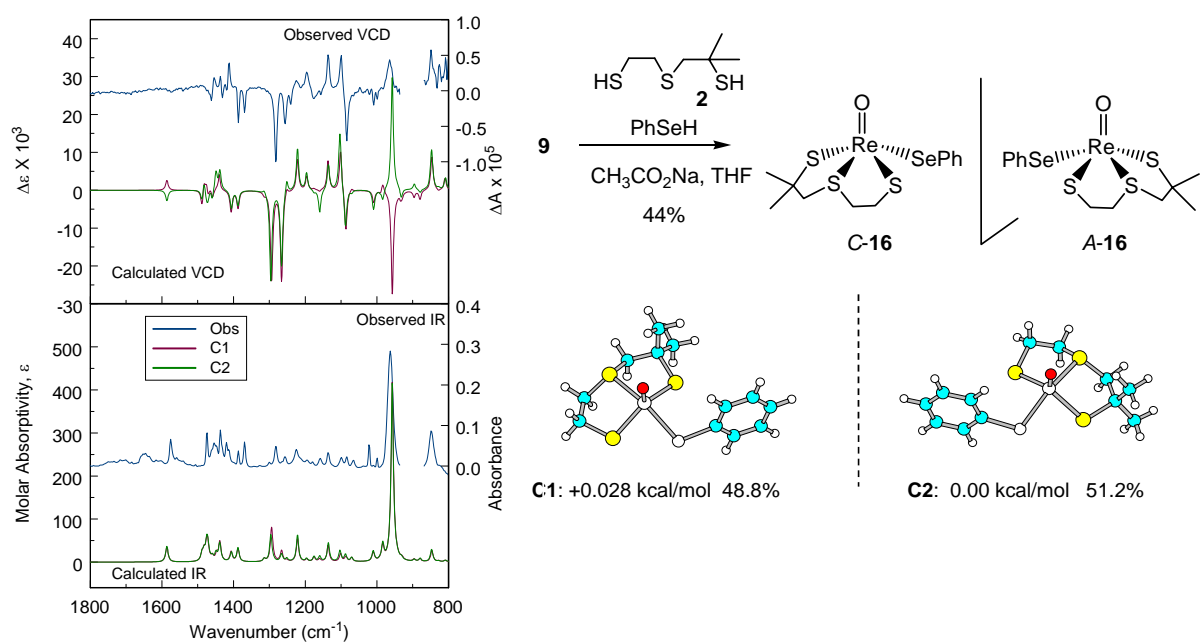

**Figure 8:** Optimized geometries, relative energies and Boltzmann populations (23°C) for conformers C1 and C2 of complex ECD$_{220}$(+)-**16**. Comparison of infrared (lower frame) and VCD (upper frame) spectra observed for complex ECD$_{350}$(+)-**16** in CDCl$_3$ solvent (right axes) with calculated spectra (left axes) for the two stable conformations, C1 and C2, with geometries and *C*- ECD$_{220}$(+)-**16** absolute configuration.

In conclusion, sulfur ligand-based complexes, just like Tp ligand-based complexes present the advantage of having an intense band around 1000 cm$^{-1}$, of being synthesizable at the scale of several grams in enantioenriched form, but in addition, they are sufficiently compact to enable the theoretical determination of their PV effect (see below).

### *(ii)* **Computational studies**

We have calculated PV vibrational frequency differences for various chiral oxorhenium complexes, including **10** and **12** discussed in the preceding section. Our results (Table II) indicate PV shifts on the order of 100 mHz or higher, which require an experimental resolution beyond 10$^{-14}$ for their observation[64]. We believe that such a resolution is attainable with the proposed high resolution laser experiment. However, it is equally clear that the predicted PV shifts are about one order of magnitude smaller than the corresponding shifts calculated for the previously proposed, but unstable compounds **5** and **6**. This can be attributed to the fact that chirality is induced only in the second sphere of coordination around the rhenium atom in compounds **10** and **12.** We therefore currently look for candidate molecules with more pronounced dissymmetry at the stereogenic center, in particular chiral derivatives of methyltrioxorhenium, to be discussed in section VIII. Another striking feature of Table II is that density functional theory (DFT) tends to predict PV vibrational frequency differences that are easily one order of magnitude smaller than the estimates provided by Hartree-Fock (HF). Although DFT is presumably the better method, the unsystematic nature of current approximate density functionals urges for calibration studies using high-end wave function based methods, such as relativistic Coupled Cluster (CC) level[87], which we are currently preparing. Finally, Table II also shows the strong sensitivity of the calculated PV shift on the chemical environment around the stereogenic center. We observe for instance that replacing an ethyl group of the synthesized complex **10** by a methyl group to give the theoretical compound **10b** reduces the PV vibrational frequency difference, calculated at the relativistic HF level, from -1.585 Hz to -1.021 Hz. We are therefore currently developing analysis tools for a better understanding of parity violation in molecules, as discussed in section V.

**Table II:** Calculated PV vibrational shifts between the two enantiomers (*A-C* or *R-S*) of chiral rhenium and osmium complexes discussed in this paper. HF : Hartree-Fock ; DFT : Density Functional Theory (B3LYP).

| Compound | Transition | Transition frequency ($cm^{-1}$) | PV shift (Hz) | |
|---|---|---|---|---|
| | | | HF | DFT |
| **5** | 0→1 | 884 | -3.085 | -1.152 |
| **6** | 0→1 | 1019 | -2.077 | -2.386 |
| **10** | 0→1 | 1012 | -1.585 | -0.102 |
| **10b** | 0→1 | 1012 | -1.021 | -0.084 |
| **12** | 0→1 | 1027 | +0.157 | +0.069 |

## VII. Supersonic expansion of a solid phase of molecules

As mentioned above, the molecules currently considered for the PV experiment are solids at ambient temperature, which adds a serious difficulty to the experimental set-up. Nevertheless, they sublimate in the 100°C-250°C range, which makes it possible to prepare them in the gas phase by simply heating the container and all the gas pipes all the way to the nozzle exit in order to prevent recondensation of the compound before the supersonic expansion. The vapor obtained can thus be seeded in carrier gas (which will impose velocity and level population distribution to the studied molecules) to form a supersonic beam. For that purpose, heating devices enabling the production of rare gas-seeded molecular beams from sublimated molecules were developed by physicists and spectroscopists of our consortium and tested on achiral solid substances (see below).

### A. Tests on achiral urethane

The three teams repectively led by T. Huet, P. Asselin and C. Chardonnet succeeded in bringing molecules from a solid phase into a supersonic beam using urethane ($H_5C_2OC(=O)NH_2$, **17**) as a test solid state molecule. With this molecule, the two spectroscopy teams actually demonstrated the feasibility of high resolution supersonic beam spectroscopy using ohmic heating devices[88]. The group of T. Huet provided spectroscopic data on the two most stable conformers of urethane using microwave and millimeterwave spectroscopy techniques, as well as quantum chemistry calculations (geometry and vibrational structure). The two conformers were observed in the 4-20 GHz range in a molecular expansion of neon ($2.5 \times 10^5$ Pa) seeded with urethane (a few percents) at a temperature of 373 K. Combined with the millimeterwave spectra recorded in a static cell (2 Pa), the pure rotation spectra gave rise to a set of rotation lines modeled by a standard Watson Hamiltonian model (rotation and centrifugal distortion parameters). *Ab initio* calculations (MP2/aug-cc-pVTZ) provided a comprehensive physical meaning of the conformational landscape. The two rotamers, only separated by a relative energy of 0.33 kJ/mol and a barrier of 3.89 kJ/mol (with ZPE corrections), present two structures differing by a rotation of about 90 degrees of the ethyl group versus the functional group. Despite the low symmetry of the less stable structure, a long range interaction makes it very stable. Additional calculations were performed with the aim to observe a small shift (a few wavenumbers) on the vibrational structure of both conformers using the FTIR technique[88]. Pierre Asselin's group recorded FTIR spectra of urethane in the 1000-1900 cm$^{-1}$ region at medium resolution (Figure 9). For these experiments, the jet-cooled conditions were not compatible with an efficient conformational relaxation. Consequently it is expected to observe two urethane conformers separated by a small energy difference of about 4 kJ/mol. Small differences between predicted frequencies of both conformers made it difficult to prove their dual presence unambiguously: band assignments have been realized with the help of theoretical calculations of both harmonic frequencies and band intensities. In several cases thorough band contour simulations have been performed in order to determine whether broad and unstructured absorption features could result from a single vibrational transition or from the overlapping of two nearby vibrations[88].

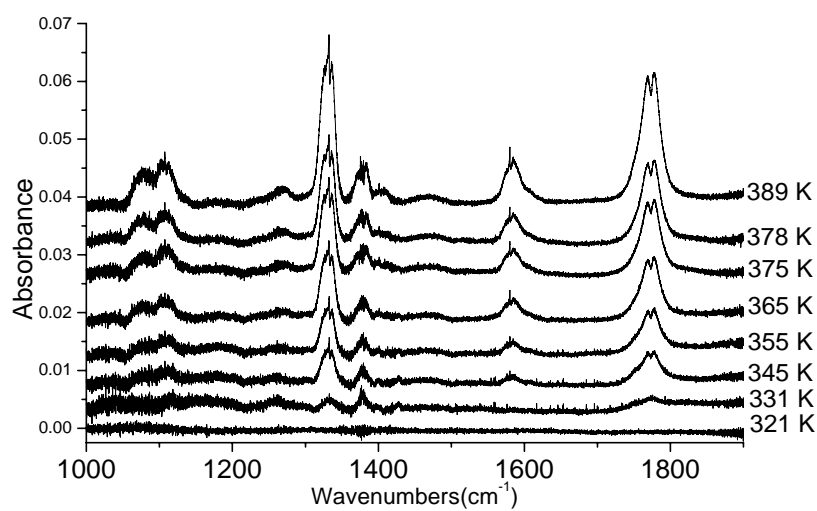

**Figure 9:** Series of jet-FTIR spectra of urethane 17/Ar mixtures recorded at 0.1 cm$^{-1}$ resolution for different temperatures of the crystalline urethane powder.

## *B.  Sublimation properties of chiral oxorhenium complexes*

Because Tp ligand-based complex **8** (section VI.B.a)) undergoes sublimation at 160°C under reduced pressure (0.2-0.3 Torr), without any decomposition, a study of its rovibrational spectrum in a molecular beam was attempted. Characterization by microwave and infrared spectroscopy has been attempted but a major drawback has been observed in the course of infrared experiments: sublimation temperatures above 250 °C, close to the limits of our ohmic heating devices, are necessary in order to obtain a ~10 mbar vapor pressure and a gas phase seeded supersonic beam with molecular flux intensities compatible with FTIR spectroscopy. At high sublimation temperature, jet-cooled molecules tend to recondense on all inner surfaces of the vacuum chamber, particularly on the optics. As a consequence, only broad and intense absorptions belonging to a solid phase of complex **8** are observed which completely hide the potential presence of weaker and narrower gas phase absorptions which fall in the same spectral region. Jet-cooled rovibrational data thus failed to be obtained for this complex.

Sublimation tests done on sulfur ligand-based oxorhenium complexes **10, 11, 12** and **14** (section VI.B.b)) have evidenced a slow decomposition between 150°C and 190°C, confirmed by NMR analysis of sublimated complexes after heating. One exception concerns the oxorhenium complex **13** bearing a SPh-4-Br for which a detectable vapor pressure, still very low compared to the Tp ligand-based complex **8,** has been observed above 195°C. These complexes are thus not suitable for a molecular beam experiment because either they tend to decompose upon sublimation or the molecular flux generated is too weak for supersonic beam spectroscopy.

## VIII. The new approach of chiral derivatives of methyltrioxorhenium (MTO)

In the light of the previous results, it seems necessary to look for other oxorhenium complexes characterized by a strong stability upon heating (as opposed to sulfur ligand-based complexes) and low sublimation temperatures of the order of 100°C (unlike Tp ligand-based complexes but like urethane). This will enable us to avoid problems with decomposition at high temperatures as well as risk of recondensation on the inner walls of vacuum chambers. Besides, compared to sulfur ligand-based complexes we look for candidate molecules with more pronounced dissymmetry around the stereogenic rhenium centre. In particular, we foresee the synthesis of better candidate molecules starting from achiral methyltrioxorhenium (MTO, Figure 10), a molecule employed in catalysis[89], since MTO is known to sublime very efficiently[90].

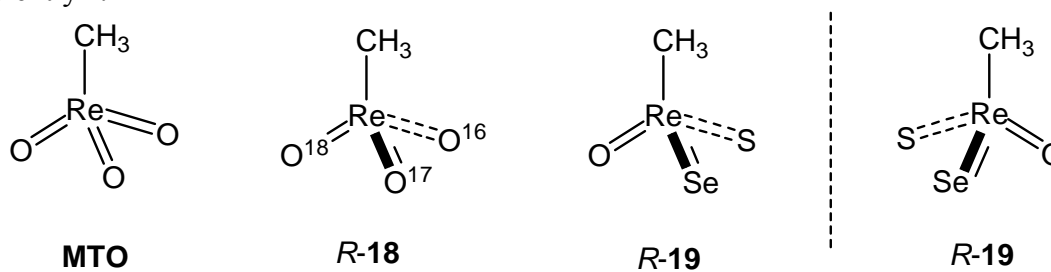

**Figure 10:** Model chiral analogues of methyltrioxorhenium MTO.

Isotopically chiral MTO ($CH_3Re^{16}O^{17}O^{18}O$, **18**, Figure 10) has been suggested by M. Quack[91], but the PV effect is expected to be small in the case of isotopic chirality. One of the ideal simple chiral molecules for a PV experiment test would be a "chiralized" MTO analogue such as $CH_3ReOSSe$ (**19**, Figure 10), that is where two oxygen atoms have been replaced by sulfur and selenium atoms. In fact, a PV vibrational frequency difference of 400 mHz has been recently estimated for this compound by T. Saue and coworkers. This value lies within the expected resolution of our experimental set-up for PV measurements. We thus want to explore the chemistry of MTO and we will focus on the development of synthesis of simple chiral derivatives of MTO. Note that chiral derivatives of MTO have already been described in literature[92-95].

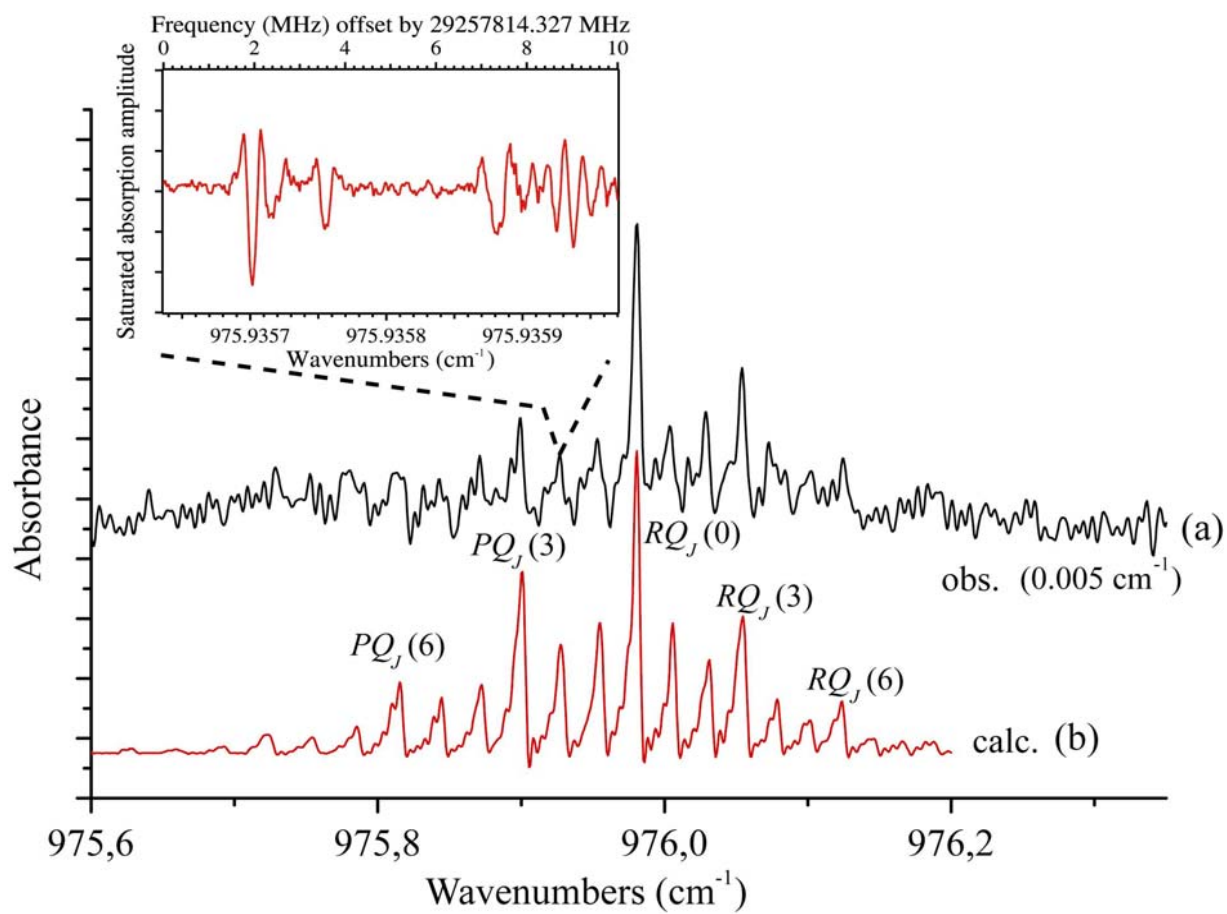

**Figure 11:** (a) measured FTIR spectrum of the $\nu_{as}$ Re=O stretching band of $CH_3{}^{187}ReO_3$; (b) stimulated spectrum (rotational temperature of 10 K) convoluted with the boxcar apodization fonction. The inset shows a saturation absorption spectrum of MTO in a cell at 300 K (detected by frequency modulation, second harmonic).

In this context, a detailed spectroscopic study of achiral MTO has been undertaken. MTO is commercially available (Strem Chemicals Inc., 98% purity) and is an ideal test molecule, being the parent molecule of potential candidates for the PV experiment. Preliminary results obtained in the groups of P. Asselin, T. Huet and C. Chardonnet are encouraging. In P. Asselin's team, jet-cooled rovibrational FTIR spectra of MTO have been successfully recorded up to 0.005 cm$^{-1}$ resolution (Figure 11) following the experimental know-how acquired with the urethane test molecule (see section VII). The best compromise between rovibrational cooling and signal intensity of the Re=O antisymmetric stretching band ($\nu_{as}$) was obtained with MTO heated at 360 K leading to a 10% dilution of MTO vapor in argon for a backing pressure of 100 mbar. With respect to experiments with chlorofluoroiodomethane **2**, the better signal-to-noise ratio of the jet-FTIR spectrum enabled to perform a more complete rovibrational analysis of the perpendicular band of this $C_{3v}$ top symmetric molecule: excited-state rovibrational parameters could be determined by a least-squares minimization between observed and synthetic spectra (Figure 11). Error bars of 100 MHz for band centers of $CH_3{}^{185}ReO_3$ and $CH_3{}^{187}ReO_3$, 300 kHz for rovibrational coupling constants and 3 % for the Coriolis parameter have been reached. The line list of the $\nu_{as}$ transitions assigned in *J* and *K* (rotational quantum numbers) obtained from this set of parameters leads to a deviation as small as 15 MHz between the 10*R*(20) line of the ultra stable $CO_2$ laser and the $PQ_2(2)$ transition of the $\nu_{as}$ band of $CH_3{}^{187}ReO_3$. This particular transition could thus be used for a realistic test of the ultimate PV experiment.

Fourier transform microwave spectra of vapors (333 K) of MTO seeded in neon ($1.6 \times 10^5$ Pa) as a carrier gas were performed in T. Huet's group. Results were compared with literature data[96]. Because of the sensitivity of their new Fourier transform microwave spectrometer (2-20 GHz range), many additional weak lines were observed, increasing by a factor two the number of reported lines. The accuracy of spectroscopic parameters was improved, especially for the hyperfine structure. For example the ratio of the experimental quadrupole constants associated with the $^{187}$Re and $^{185}$Re atoms of MTO was found to be 0.946 323 5(34) instead of 0.946 36(6), with an uncertainty improved by a factor twenty.

Simultaneously, C. Chardonnet's team recently achieved saturated absorption spectroscopy of MTO at 300 K in a 60-cm long cell with a resolution of 50 kHz. Dense rovibrational spectra with hyperfine sub-structures were observed in a 500 MHz spectral window around the *R*(20) $CO_2$ laser line at 10.2 µm (see inset on Figure 11). A supersonic beam of MTO was also detected, via time of flight experiments, on the device dedicated to the observation of PV.

These results clearly demonstrate the possibility to produce rare gas seeded supersonic molecular beams from sublimated organometallic molecules and to perform high resolution spectroscopy in the infrared and the microwave domain on them. We are therefore confident in our ability to observe and characterize new chiral derivatives of MTO.

## IX. Conclusion

In this paper, we have described the combined efforts during recent years of physicists and chemists, experimentalists and theoreticians, towards the observation of parity violation in molecular systems. This work shows the interplay between disciplines as diverse as relativistic quantum chemistry, synthetic chemistry, spectroscopy, frequency metrology and fundamental physics. This collaboration started 10 years ago when in 1999 the synthesis and separation of the enantiomers of CHFClBr by A. Collet and J. Crassous enabled a first test of parity violation by the LPL group, using saturation spectroscopy in absorption cells. This experiment having shown its limits, a new set-up based on molecular beam spectroscopy using the two-photon Ramsey fringes technique is currently being developed.

In parallel, chiral molecules such as CHFClI, Tp ligand-based and sulfur ligand based oxorhenium complexes have been prepared in enantiomerically enriched forms. Transition metal complexes such as oxorhenium complexes display more pronounced PV effects, but it has been up to now impossible to prepare a molecular beam with such molecules. It appears necessary to look for other oxorhenium complexes with higher stability upon heating and lower sublimation temperatures (of the order of 100°C). For this reason we now focus on a new class of compounds which are analogues of methyltrioxorhenium (MTO).

Concurrently, relativistic calculations have been conducted. Our theoretical studies show an extreme sensitivity of the calculated PV vibrational frequency difference with respect to the chemical environment around the stereogenic center. On the other hand, we note that this sensitivity makes the present project chemically even more *interesting*. The situation can be compared to the situation for nuclear magnetic resonance, where radio frequency spectroscopy involving nuclear spin states, seemingly remote from chemistry, evolved into one of the main techniques in chemistry for the characterization of molecules and chemical transformations. We certainly do not believe that PV measurements will follow the same historical path, but we do believe that studies of PV effects in molecules can provide deep insights about chirality in molecular systems.

One should note that the present project will impact the many domains involved in our consortium. The experiment currently under development constitutes a significant advance in the domain of laser spectroscopy and should make available ultra-high resolution spectroscopy techniques for complex molecules. Likewise, we expect the synthesis of suitable candidate molecules to engender further refinement and developments in the very active area of chiral synthesis, resolution and stereochemical characterization. A successful experiment may allow a head-on confrontation between experiment and calculations of relativistic quantum chemistry and would enable low-energy tests of the weak interaction itself[97]. This would require the theoretical chemists to resort to highly correlated molecular electronic structure methods.

A fascinating aspect of PVED in chiral molecules is the possible link to the origin of biochirality. Although the role of the weak interaction in the origin of homochiral life is subject to debate[98-104], the experimental observation of parity violation effects in molecules would certainly re-activate this debate.

## Acknowledgements

This work was supported by ANR (Contract grant number : ANR-05-BLAN-0091), Ministère de la Recherche et de l'Enseignement Supérieur, CNRS, Région Bretagne, Rennes Métropole, and the Marsden Fund (New Zealand). O. Lopez, F. de Montigny, I. Karamé, F. Monier, C. Genre, J.-P. Dutasta, J.-R. Aviles-Moreno, A. Cuisset, J. Demaison, M. Goubet, M. Rey, M. Tudorie, A. Severo Pereira Gomes are gratefully acknowledged for their contributions to this project.